\documentclass[12pt]{article}
\usepackage{amsfonts}
\usepackage{lscape}
\newcommand{\sfrac}[2]{{\textstyle{\frac{#1}{#2}}}} 
\newcommand{\half}{{\textstyle{\frac{1}{2}}}}

\begin{document}
\newpage \pagestyle{empty} \setcounter{page}{0} \vfill
\begin{center}

{\Large \textbf{Theoretical Prevision of Physical-Chemical Properties of 
Amino Acids from Genetic Code}}

\vspace{10mm}

{\large L. Frappat$^a$, A. Sciarrino$^{b}$, P. Sorba$^a$}

\vspace{10mm}

\emph{$^a$ Laboratoire d'Annecy-le-Vieux de Physique Th{\'e}orique LAPTH}

\emph{CNRS, UMR 5108, associ{\'e}e {\`a} l'Universit{\'e} de Savoie}

\emph{BP 110, F-74941 Annecy-le-Vieux Cedex, France}

\vspace{7mm}

\emph{$^b$ Dipartimento di Scienze Fisiche, Universit{\`a} di Napoli 
``Federico II''}

\emph{and I.N.F.N., Sezione di Napoli}

\emph{Complesso Universitario di Monte S. Angelo, Via Cintia, I-80126 
Napoli, Italy}

\end{center}

\vspace{12mm}

\begin{abstract}
Using the crystal basis model of the genetic code, a set of relations 
between the physical-chemical properties of the amino acids are derived and 
compared with the experimental data.  A prevision for the not yet measured 
thermodynamical parameters of three amino acids is done.
\end{abstract}

\vfill
\rightline{LAPTH-805/00}
\rightline{DSF-TH-24/00}
\rightline{physics/0007034}
\rightline{July 2000}

\clearpage
\pagestyle{plain}
\baselineskip=18pt

%%%%%%%%%%%%%%%%%%%%%%%%%%%%%%%%%%%%%%%%%%%%%%%%%%%%%%%%%%%%%%%%%%%%%%%

\section{Introduction}

It is a known observation \cite{SW} that a relationship exists between the 
codons and the physical-chemical properties (PCP) of the coded amino acids 
(a.a.).  The observed pattern is read either as a relic of some kind of 
interaction between the a.a. and the nucleotides at an early stage of 
evolution or as the existence of a mechanism relating the properties of 
codons with those of a.a..  In particular it is observed that the 
relationship depends essentially on the nature of the \emph{second} 
nucleotide in the codons and it holds when the second nucleotide is adenine 
(A), uracil (U) or cytosine (C), not when it is guanine (G).  To our 
knowledge neither the anomalous behaviour of G nor the existence of a 
closest relationship between some of the a.a. is understood.  It is the aim 
of this paper to provide an explanation of both these facts in the 
framework of the model of genetic code we have discussed in a previous 
paper \cite{FSS1}, called crystal basis model, as it is heavily based on 
the mathematical structure of the modules of the irreducible 
representations (IR) of ${\cal U}_{q \to 0}[sl(2) \oplus sl(2)]$, known in 
the mathematical literature as \emph{crystal basis} \cite{Kashi}.  After 
recalling briefly the model in Sect.  2, we derive in Sect.  3 a set of 
relations between the PCP of the amino acids, based on the content of the 
IRs of the dinucleotides and of the codons coding for the amino acids.  
Finally we compare our predictions with the experimental data.

\section{The Model}

The four nucleotides A, C, G, U are considered as basic states of the 
$(\half,\half)$ irreducible representation of the quantum enveloping 
algebra ${\cal U}_{q}[sl(2) \oplus sl(2)]$ in the limit $q \to 0$, with the 
following assignment of quantum numbers:
\begin{equation}
	\mbox{C} \equiv (+\half,+\half) \qquad \mbox{U} \equiv (-\half,+\half) 
	\qquad \mbox{G} \equiv (+\half,-\half) \qquad \mbox{A} \equiv 
	(-\half,-\half)
	\label{eq:gc1}
\end{equation}
In this framework, the codons are built as composite states of the 
nucleotide states by tensoring three such $(\half,\half)$ representations.  
Note that the crystal basis, which exists in the limit $q \to 0$ of the 
$q$-deformed universal enveloping algebra $U_q(G)$ for any semi-simple Lie 
algebra $G$, is the only way to provide such composite states as 
\emph{pure} states, and to ensure the existence of an \emph{order} in the 
tensoring procedure.

However, it is well-known (see Table \ref{tablerep}) that in a multiplet of 
codons relative to a specific amino acid, the first two bases constituent 
of a codon are ``relatively stable'', the degeneracy being mainly generated 
by the third nucleotide.  Considering the tensor product
\begin{equation}
	(\half,\half) \, \otimes \, (\half,\half) = (1,1) \, \oplus \, (1,0) \, 
	\oplus \, (0,1) \, \oplus \, (0,0)
	\label{eq:gc4}
\end{equation}
we get, using Kashiwara's theorem \cite{Kashi}, the following tableau:
\begin{displaymath}
	\begin{array}{lcccc}
		\to \,\, sl(2)_H \qquad \qquad &(0,0) & (\mbox{CA}) &\qquad 
		\qquad (1,0) & (
		\begin{array}{ccc} 
			\mbox{CG} & \mbox{UG} & \mbox{UA} \\
		\end{array}) \\
		\downarrow \\
		sl(2)_V &(0,1) & \left(
		\begin{array}{c}
			\mbox{CU} \\
			\mbox{GU} \\
			\mbox{GA} \\
		\end{array}
		\right) & \qquad \qquad (1,1) & \left(
		\begin{array}{ccc}
			\mbox{CC} & \mbox{UC} & \mbox{UU} \\
			\mbox{GC} & \mbox{AC} & \mbox{AU} \\
			\mbox{GG} & \mbox{AG} & \mbox{AA} \\
		\end{array}
		\right)
	\end{array}
	\label{eq:dinucl}
\end{displaymath}
where the subscripts $H$ (:= horizontal) and $V$ (:= vertical) specify the 
two $sl(2)$.

{From} Table \ref{tablerep}, the dinucleotide states formed by the first 
two nucleotides in a codon can be put in correspondence with quadruplets, 
doublets or singlets of codons relative to an amino acid, the sextets 
(resp.  triplets) being viewed as the sum of a quadruplet and a doublet 
(resp.  a doublet and a singlet).  We define in our model the ``charge'' 
$Q$ of a dinucleotide state by
\begin{equation}
	Q = J_{3,H} + \sfrac{1}{4} \, C_{V} (J_{3,V}+1) - \sfrac{1}{4}
	\label{eq:charge}
\end{equation}
The dinucleotide states are then split into two octets with respect to the 
charge $Q$: the eight \emph{strong} dinucleotides CC, GC, CG, GG, CU, GU, 
UC, AC associated to the quadruplets (as well as those included in the 
sextets) of codons satisfy $Q > 0$, while the eight \emph{weak} 
dinucleotides AA, AU, UA, UU, UG, AG, CA, GA associated to the doublets (as 
well as those included in the triplets) and eventually to the singlets of 
codons satisfy $Q < 0$.  Let us remark that by the change C 
$\leftrightarrow$ A and U $\leftrightarrow$ G, which is equivalent to the 
change of the sign of $J_{3, \alpha}$ or to reflexion with respect to the 
diagonals of the eq.  (\ref{eq:gc1}), the 8 strong dinucleotides are 
transformed into weak ones and vice-versa.  Note that a first attempt to 
differentiate between strong and weak dinucleotides is contained in ref.  
\cite{KR}.

The irreducible representations of the tensor product of $(\half, 
\half)^{\otimes 3}$ as well as the correspondence codons/amino acids for 
the eukariotic code is reported in Table \ref{tablerep}.  The upper labels 
denote different IRs.

\section{Relationship between the PCP of Amino Acids}
 
We assume that some PCP of a given amino acid are related to the nature of 
the codons, in particular they depend on the following mathematical 
features, written in hierarchical order:
\begin{enumerate}
	\item
	the IR of the dinucleotide formed by the first two nucleotides;
	\item
	the sign of the charge $Q$, eq.  (\ref{eq:charge}), on the dinucleotide 
	state;
	\item
	the value of the third component of $J_{3,V}$ inside a fixed IR for the 
	dinucleotides;
	\item
	the upper label(s) of the codon IR(s);
\end{enumerate}
Not all the PCP are supposed to follow the scheme above; some of them are 
essentially given by the specific chemical structure of the amino acid 
itself.

\medskip

In the following, we analyze the PCP of the amino acids in the light of the 
dinucleotide content of the irreducible representations of eq.  
(\ref{eq:dinucl}).  

\medskip

-- representation (0,0) \\
The codons of the form CAN (N = C, U, G, A) all belong to the IR $(\half, 
\half)^{4}$ and code for \texttt{His} and \texttt{Gln}, both being coded by 
doublets and differing by the value of $J_{3,V}$.  Then we expect that the 
PCP of \texttt{His} and \texttt{Gln} are very close.  

\medskip

-- representation (1,0) \\
We analyse now the codons built up by the dinucleotide IR ($1,0$).  i.e.  
CG ($Q > 0$), UG, UA (both $Q < 0$).  The codons CGS (S = C, G), resp.  CGW 
(W = U, A), belonging to IR $(3/2, 1/2)^{2}$, resp.  $(1/2, 1/2)^{2}$, all 
code for \texttt{Arg}, so we do not have any relation.  The codons UGS, 
resp.  UGW, belonging to IR $(3/2, 1/2)^{2}$, resp.  $(1/2, 1/2)^{2}$, code 
for \texttt{Cys} (which is a doublet) and \texttt{Trp} (singlet), resp.  
the other \texttt{Cys} and \texttt{Ter} (triplets).  So we expect the PCP 
of \texttt{Cys} not very different from those of \texttt{Trp}.  The codons 
UAN, belonging to the IR $(3/2,1/2)^{2}$, code for the \texttt{Tyr} and 
\texttt{Ter}.  So we expect some affinity between the a.a.  coded by UGN 
and UAN, in particular between \texttt{Cys} and \texttt{Tyr} both being 
coded by doublets.  

\medskip

-- representation (0,1) \\
The codons built up by the dinucleotide IR ($0,1$) are CU, GU (both $Q > 
0$) and GA ($Q < 0$).  The codons CUY and GUY (Y = C, U), resp.  CUR and 
GUR (R = G, A), belonging to IR $(1/2, 3/2)^{2}$, resp.  $(1/2, 1/2)^{3}$, 
code for \texttt{Leu} and \texttt{Val}.  Therefore we do not have any 
relation between a.a.  coded by the same dinucleotide, but we expect that 
the PCP of \texttt{Leu} and \texttt{Val} are close since CU and GU both 
belong to the same IR and are both strong.  The codons GAN belong to the IR 
$(1/2, 3/2)^{2}$ and they code \texttt{Asp} and \texttt{Glu} (both 
doublets).  Then we expect the PCP of Asp and Glu to be very close.  

\medskip

-- representation (1,1) \\
The dinucleotide IR ($1,1$) contains five states with $Q > 0$ (CC, UC, GC, 
AC, GG).  The codons CCN , resp.  UCN, belong to four different IRs and 
code for \texttt{Pro} (quartet), resp.  \texttt{Ser} (sextets).  We expect 
an affinity between the PCP of these a.a..  The codons GCN , resp.  ACN, 
belong to four different IRs and code for \texttt{Ala} (quartet), resp.  
\texttt{Thr} (quartets).  We expect a strong affinity between the PCP of 
\texttt{Ala} and \texttt{Thr}.  The codons GGN belong to two different IRs 
and code for \texttt{Gly}, so we expect an affinity of PCP of \texttt{Gly} 
with those of \texttt{Pro}, \texttt{Ser}, \texttt{Ala}, \texttt{Thr}.  Now 
let us look at the four states with $Q < 0$ (UU, AU, AG, AA).  The codons 
UUN belong to two different IRs and code for \texttt{Leu}, the doublet 
subpart of the sextet, and for \texttt{Phe} (doublet).  An affinity is 
expected between the PCP of these two a.a..  The codons AUN belong to two 
different IRs and code \texttt{Ile} (triplet) and \texttt{Met} (singlet) 
and, in fact, the values of PCP of these two a.a.  are not very different.  
The codons AGN belong to two different IRs and code for \texttt{Ser} and 
\texttt{Arg}, the doublet subpart of the sextet, so an affinity between the 
PCP of these codons is expected.  The codons AAN belong to the same IR 
($3/2, 3/2$) and code for \texttt{Asn} and \texttt{Lys}, so the values of 
the PCP of these a.a.  should be close.

\medskip

Note that for the three sextets (\texttt{Arg}, \texttt{Leu}, \texttt{Ser}) 
the quartet (doublet) subpart is coded by a codon with a strong (weak) 
dinucleotide.

In conclusion we predict the following relations between the values of the 
PCP of the amino acids ($\cong$ means strong affinity, $\approx$ affinity, 
$\sim$ weak affinity):
\begin{enumerate}
	\item
	His $\cong$ Gln
	\item
	Asp $\cong$ Glu
	\item
	Asn $\cong$ Lys $\sim$ Arg, Ser
	\item
	Cys $\cong$ Tyr $\approx$ Trp
	\item
	Leu $\cong$ Val
	\item
	Pro $\cong$ Ser $\approx$ Gly
	\item
	Ala $\cong$ Thr $\approx$ Gly, Pro, Ser
	\item
	Ile $\cong$ Met $\approx$ Phe
\end{enumerate}

\section{Discussion}

We have compared our theoretical previsions with 10 physical-chemical 
properties:
\begin{enumerate}
	\item
	the Chou-Fasman conformational parameters \cite{CF} $P_{\alpha}$, 
	$P_{\beta}$ and $P_{\tau}$ which gives a measure of the probability of 
	the a.a. to form respectively a helix, a sheet and a turn.  However it 
	has been suggested in \cite{S33} that the sum $P_{\alpha} + P_{\beta}$ 
	is a more appropriate parameter to characterize the generic structure 
	forming potential while the difference $P_{\alpha} - P_{\beta}$ is a 
	more appropriate parameter for the helix forming potential, which is a 
	quantity more depending on the particular a.a..  So we compare with 
	$P_{\alpha}$ + $P_{\beta}$ and $P_{\tau}$;
	\item
	the Grantham polarity $P_G$ \cite{G};
	\item
	the relative hydrophilicity $R_f$ as computed by Weber and Lacey 
	\cite{WL};
	\item
	the thermodynamic activation parameters at 298 K: $\Delta H$ (enthalpy, 
	in kJ/mol), $\Delta G$ (free energy, in kJ/mol) and $\Delta S$ 
	(entropy, in J/mole/K) as obtained by Siemion and Stefanowicz 
	\cite{SS};
	\item
	the negative of the logarithm of the dissociation constants at 298 K: 
	$pK_a$ for the $\alpha$-COOH group and $pK_b$ for the 
	$\alpha$-NH$_3^{+}$ group \cite{Hand};
	\item
	the isoelectronic point $pI$ \cite{Hand}, i.e. the $pH$ value at which 
	no electrophoresis occurs.
\end{enumerate}
The comparison between the theoretical relations and the experimental 
values shows (see Tables \ref{tableTE} and \ref{tableSD}):
\begin{enumerate}
	\item
	His $\cong$ Gln -- The agreement, except for $pI$, is very good.
	\item
	Asp $\cong$ Glu -- The agreement, except for $P_{\tau}$, is very good.
	\item
	Asn $\cong$ Lys $\sim$ Arg, Ser -- The agreement, except for $pI$ and 
	$P_{\tau}$ is very good.  The comparison with the values of PCP of Ser 
	and Arg is satisfactory.
	\item
	Cys $\cong$ Tyr $\approx$ Trp -- Except for $R_f$, the agreement 
	between the first two a.a. is very good, while with Trp is 
	satisfactory.
	\item
	Leu $\cong$ Val -- The agreement is very good.
	\item
	Pro $\cong$ Ser $\approx$ Gly -- The agreement is very good, except for 
	$P_{\alpha} + P_{\beta}$ and $\Delta H$, and with Gly more than 
	satisfactory.
	\item
	Ala $\cong$ Thr $\approx$ Gly, Pro, Ser -- The agreement is very good 
	between the first two a.a. except for $P_{\tau}$ and satisfactory with 
	the others except for the conformational parameters.
	\item
	Ile $\cong$ Met $\approx$ Phe -- The agreement is very good between the 
	first two a.a. and satisfactory with Phe.
\end{enumerate}
In order to have a more quantitative evaluation of the agreement between 
the data and the theoretical model we compute the mean value and the 
standard deviation of the whole population (i.e. the twenty amino-acids) 
for each generic PCP $x$
\begin{equation}
	\overline{x} = \frac{1}{n} \, \sum_i^{n} \, x_i \quad \mbox{and} \quad 
	\sigma_x = \sqrt{\frac{1}{n} \, \sum_i^{n} \, (x_i - \overline{x})^2}
\end{equation}
in 3 different cases:
\begin{enumerate}
	\item
	considering all the amino acids, i.e. summing over the whole set of 
	amino acids ($n$ = 20);
	\item
	considering the amino acids coded by the codons with the same second 
	nucleotide ($n$ = 4 for C, $n$ = 5 for U and G, $n$ = 7 for A);
	\item
	considering the couples ($n$ = 2) of amino acids given by our 8 
	relations.
\end{enumerate}
The results are reported in Tables \ref{tableTE} to \ref{tableSD}.
\\
As an estimate of the accurary of our predictions, we compute the sum of 
the adimensional quantities given for any PCP by the ratios 
$\sigma_{x}/\overline{x}$ of the standard deviation to the average value, 
in the three different cases considered above.
\begin{center}
	\begin{tabular}{|c|c|c|c|c|c|c|c|}
		\hline
		\multicolumn{8}{|c|}{all \qquad 1.82} \\
		\hline
		\multicolumn{2}{|c|}{A \qquad 1.82} & \multicolumn{2}{|c|}{C \qquad 
		1.13} & \multicolumn{2}{|c|}{G \qquad 1.82} & 
		\multicolumn{2}{|c|}{U \qquad 0.70} \\
		\hline
		Gln/His & Asp/Glu & Asn/Lys & Cys/Tyr & Pro/Ser & Ala/Thr & Ile/Met 
		& Leu/Val \\
		0.35 & 0.63 & 0.72 & 0.80 & 0.77 & 0.48 & 0.39 & 0.38 \\
		\hline
	\end{tabular}
	
	\medskip
	
	Values of $\eta = \sum \sigma_{x}/\overline{x}$
\end{center}
Looking to the table above, we remark that the characterization of the PCP 
from the nature of the second nucleotide in the codons is, except for U, 
indeed not really discriminatory.  On the contrary, the relations derived 
by our model reduces dramatically the values of our estimate parameter 
$\eta = \sum_{i} \sigma_{x}/\overline{x}$.

\medskip

{From} Table \ref{tableTE} we note that the values of the thermodynamical 
quantities for some amino acids are not yet measured.  Our relations 
predict that the values of $\Delta H$ and $-\Delta S$ for \texttt{His} 
should be around 60 kJ/mol and 120 kJ/mol/K. For the values of \texttt{Asp} 
and \texttt{Glu} coded by the unique weak nucleotide GA in the IR (0,1), we 
do not have any available data to compare with.  However, making the 
further assumption, supported by an inspection of the available data, that 
the values of the thermodynamical quantities depend slightly on the nature 
of the charge, we predict that the values of these quantities for 
\texttt{Asp} and \texttt{Glu} should not be very dissimilar from the values 
of the analogous quantities for \texttt{Leu} and \texttt{Val}, coded by the 
strong dinucleotide CU an GU in the IR (0,1).  So we predict that for 
\texttt{Asp} and \texttt{Glu}, one should find $\Delta H \approx 60$ 
kJ/mol, $-\Delta S \approx 135$ kJ/mol/K and $\Delta G \approx 100$ kJ/mol.

\section{Conlusion}
 
In conclusion, the values of PCP show, with a few exceptions, a pattern of 
correlations which is expected from the assumptions done in the crystal 
basis model.  Let us emphasize that the assignment of the value of the IRs 
both for the dinucleotide and for the trinucleotide states or codons is a 
straightforward consequence of our model.  The remarked property that the 
a.a. coded by codons whose second nucleotide is G do not share similarity 
in the properties of PCP with other a.a. does find an explication in the 
model as it is immediate to verify, looking to the table of the tensor 
product of two ($\half, \half$) that that are no two states with G in 
second position which share simultaneously the properties of belonging to 
the same IR and being characterized by the same value of $Q$.

\medskip

A final remark: it has beeen sugested that the physico-chemical properties 
of the amino acids have played a fundamental role in the evolution and the 
organization of the genetic code.  In this framework, a meaure of the 
relative distances between amino acids has been defined, see \cite{DG} and 
references therein.  The picture which emerges in \cite{DG} has some 
striking confirmations with the predictions of our model.  So we argue that 
the assignment of codons in multiplets, corresponding to irreducible 
representations, may have a deep connection with the evolutionary 
organization.

\bigskip
 
\textbf{Acknowledgments:} We are grateful to M.L. Chiusano for pointing us 
the interest to analyse the physical-chemical properties in the light of 
the crystal basis model and for providing us literature on the subject.  
Partially supported by MURST (Italy) and MAE  (France) in the framework of 
french-italian collaboration Galileo.

\clearpage

\begin{table}[t]
	\caption{The eukariotic code.  The upper label denotes different 
	irreducible representations.}
	\label{tablerep}
	\footnotesize
	\begin{center}
		\begin{tabular}{|cc|cc|cc|cc|}
			\hline
			codon & a.a. & $J_{H}$ & $J_{V}$ & codon & a.a. & $J_{H}$ & 
			$J_{V}$ \\
			\hline
			CCC & Pro & 3/2 & 3/2 & UCC & Ser & 3/2 & 3/2 \\
			CCU & Pro & (1/2 & 3/2)$^1$ & UCU & Ser & (1/2 & 3/2)$^1$ \\
			CCG & Pro & (3/2 & 1/2)$^1$ & UCG & Ser & (3/2 & 1/2)$^1$ \\
			CCA & Pro & (1/2 & 1/2)$^1$ & UCA & Ser & (1/2 & 1/2)$^1$ \\
			\hline
			CUC & Leu & (1/2 & 3/2)$^2$ & UUC & Phe & 3/2 & 3/2 \\
			CUU & Leu & (1/2 & 3/2)$^2$ & UUU & Phe & 3/2 & 3/2 \\
			CUG & Leu & (1/2 & 1/2)$^3$ & UUG & Leu & (3/2 & 1/2)$^1$ \\
			CUA & Leu & (1/2 & 1/2)$^3$ & UUA & Leu & (3/2 & 1/2)$^1$ \\
			\hline
			CGC & Arg & (3/2 & 1/2)$^2$ & UGC & Cys & (3/2 & 1/2)$^2$ \\
			CGU & Arg & (1/2 & 1/2)$^2$ & UGU & Cys & (1/2 & 1/2)$^2$ \\
			CGG & Arg & (3/2 & 1/2)$^2$ & UGG & Trp & (3/2 & 1/2)$^2$ \\
			CGA & Arg & (1/2 & 1/2)$^2$ & UGA & Ter & (1/2 & 1/2)$^2$ \\
			\hline
			CAC & His & (1/2 & 1/2)$^4$ & UAC & Tyr & (3/2 & 1/2)$^2$ \\
			CAU & His & (1/2 & 1/2)$^4$ & UAU & Tyr & (3/2 & 1/2)$^2$ \\
			CAG & Gln & (1/2 & 1/2)$^4$ & UAG & Ter & (3/2 & 1/2)$^2$ \\
			CAA & Gln & (1/2 & 1/2)$^4$ & UAA & Ter & (3/2 & 1/2)$^2$ \\
			\hline
			GCC & Ala & 3/2 & 3/2 & ACC & Thr & 3/2 & 3/2 \\
			GCU & Ala & (1/2 & 3/2)$^1$ & ACU & Thr & (1/2 & 3/2)$^1$ \\
			GCG & Ala & (3/2 & 1/2)$^1$ & ACG & Thr & (3/2 & 1/2)$^1$ \\
			GCA & Ala & (1/2 & 1/2)$^1$ & ACA & Thr & (1/2 & 1/2)$^1$ \\
			\hline
			GUC & Val & (1/2 & 3/2)$^2$ & AUC & Ile & 3/2 & 3/2 \\
			GUU & Val & (1/2 & 3/2)$^2$ & AUU & Ile & 3/2 & 3/2 \\
			GUG & Val & (1/2 & 1/2)$^3$ & AUG & Met & (3/2 & 1/2)$^1$ \\
			GUA & Val & (1/2 & 1/2)$^3$ & AUA & Ile & (3/2 & 1/2)$^1$ \\
			\hline
			GGC & Gly & 3/2 & 3/2 & AGC & Ser & 3/2 & 3/2 \\
			GGU & Gly & (1/2 & 3/2)$^1$ & AGU & Ser & (1/2 & 3/2)$^1$ \\
			GGG & Gly & 3/2 & 3/2 & AGG & Arg & 3/2 & 3/2 \\
			GGA & Gly & (1/2 & 3/2)$^1$ & AGA & Arg & (1/2 & 3/2)$^1$ \\
			\hline
			GAC & Asp & (1/2 & 3/2)$^2$ & AAC & Asn & 3/2 & 3/2 \\
			GAU & Asp & (1/2 & 3/2)$^2$ & AAU & Asn & 3/2 & 3/2 \\
			GAG & Glu & (1/2 & 3/2)$^2$ & AAG & Lys & 3/2 & 3/2 \\
			GAA & Glu & (1/2 & 3/2)$^2$ & AAA & Lys & 3/2 & 3/2 \\
			\hline
		\end{tabular}
	\end{center}
\end{table}

\clearpage

\begin{table}
	\footnotesize
	\centering
	\begin{tabular}{|c||c|c|c|c|c|c|c|c|c|c|}
		\hline
		Amino acid & $P_{\alpha} + P_{\beta}$ & $P_{\tau}$ & $P_{G}$ & 
		$R_{f}$ & $\Delta H$ & $-\Delta S$ & $\Delta G$ & $ pK_{a}$ & 
		$pK_{b}$ & $pI$ \\
		\hline \hline
		Ala & 2.25 & 0.66 & 8.09 & 0.89 & 50.7 & 147.8 & 94.66 & 2.34 & 
		9.69 & 6.00 \\
		Arg & 1.91 & 0.95 & 10.50 & 0.88 & 54.8 & 143.2 & 97.5 & 2.17 & 
		9.04 & 11.15 \\
		Asn & 1.60 & 1.56 &11.5 & 0.89 & 55 & 135.8 & 95.51 & 2.02 & 8.80 & 
		5.41 \\
		Asp & 1.55 & 1.46 & 13 & 0.87 & $-$ & $-$ & $-$ & 1.88 & 9.60 & 
		2.77 \\
		Cys & 1.89 & 1.19 & 5.5 & 0.85 & 43.9 & 163.2 & 92.89 & 1.96 & 
		10.28 & 5.02 \\
		Gln & 2.21 & 0.98 & 10.5 & 0.82 & 61.1 & 121.8 & 97.30 & 2.17 & 
		9.13 & 5.65 \\
		Glu & 1.88 & 0.74 & 12.2 & 0.84 & $-$ & $-$ & $-$ & 2.19 & 9.67 & 
		3.22 \\
		Gly & 1.32 & 1.56 & 9 & 0.92 & 49.0 & 140.6 & 91.08 & 2.34 & 9.60 & 
		5.97 \\
		His & 1.87 & 0.95 & 10.4 & 0.83 & $-$ & $-$ & 94.59 & 1.82 & 9.17 & 
		7.47 \\
		Ile & 2.68 & 0.47 & 5.2 & 0.76 & 57.3 & 152.3 & 102.6 & 2.36 & 9.60 
		& 5.94 \\
		Leu & 2.51 & 0.59 & 4.9 & 0.73 & 59.4 & 128.1 & 97.70 & 2.36 & 9.60 
		& 5.98 \\
		Lys & 1.90 & 1.01 & 11.3 & 0.97 & 57.8 & 135.6 & 98.16 & 2.18 & 
		8.95 & 9.59 \\
		Met & 2.50 & 0.60 & 5.7 & 0.74 & 58.2 & 128.5 & 96.65 & 2.28 & 9.21 
		& 5.74 \\
		Phe & 2.51 & 0.60 & 5.2 & 0.52 & 48.3 & 156.1 & 94.76 & 1.83 & 9.13 
		& 5.48 \\
		Pro & 1.12 & 1.52 & 8 & 0.82 & 50.7 & 164.5 & 99.49 & 1.99 & 10.60 
		& 6.30 \\
		Ser & 1.52 & 1.43 & 9.19 & 0.96 & 36.0 & 180 & 89.98 & 2.21 & 9.15 
		& 5.68 \\
		Thr & 2.02 & 0.96 & 8.59 & 0.92 & 53.4 & 136.0 & 93.98 & 2.09 & 
		9.10 & 5.64 \\
		Trp & 2.45 & 0.96 & 5.4 & 0.2 & 51.1 & 151.5 & 96.25 & 2.83 & 9.39 
		& 5.89 \\
		Tyr & 2.16 & 1.14 & 6.2 & 0.49 & 54.4 & 139.0 & 95.68 & 2.20 & 9.11 
		& 5.66 \\
		Val & 2.76 & 0.50 & 5.9 & 0.85 & 60.2 & 139.0 & 101.7 & 2.32 & 9.62 
		& 5.96 \\
		\hline \hline
		$\overline{x}$ & 2.03 & 0.99 & 8.31 & 0.79 & 53.02 & 144.88 & 96.14 
		& 2.18 & 9.42 & 6.03 \\
		$\sigma_{x}$ & 0.45 & 0.36 & 2.61 & 0.18 & 6.23 & 15.57 & 3.16 & 
		0.23 & 0.43 & 1.76 \\
		$\sigma_{x}/\overline{x}$ & 22 \% & 36 \% & 31 \% & 23 \% & 12 \% & 
		10 \% & 3 \% & 10 \% & 5 \% & 29 \% \\
		\hline
	\end{tabular}
	\caption{Table of amino-acid properties} \label{tableTE}
\end{table}

% \clearpage

\begin{table}
	\footnotesize
	\centering
	\begin{tabular}{l|c||c|c|c|c|c|c|c|c|c|c|}
		\cline{2-12} & & $P_{\alpha} + P_{\beta}$ & $P_{\tau}$ & $P_{G}$ & 
		$R_{f}$ & $\Delta H$ & $-\Delta S$ & $\Delta G$ & $ pK_{a}$ & 
		$pK_{b}$ & $pI$ \\
		\cline{2-12} \cline{2-12} & $\overline{x}$ & 1.88 & 1.12 & 10.73 & 
		0.82 & 57.08 & 133.05 & 96.25 & 2.07 & 9.20 & 5.68 \\
		A & $\sigma_{x}$ & 0.23 & 0.27 & 2.03 & 0.14 & 2.65 & 6.63 & 1.29 & 
		0.15 & 0.30 & 2.17 \\
		& $\sigma_{x}/\overline{x}$ & 12 \% & 24 \% & 19 \% & 17 \% & 5 \% 
		& 5 \% & 1 \% & 7 \% & 3 \% & 38 \% \\
		\cline{2-12} \cline{2-12} & $\overline{x}$ & 1.73 & 1.14 & 8.47 & 
		0.90 & 47.70 & 157.08 & 94.53 & 2.16 & 9.64 & 5.90 \\
		C & $\sigma_{x}$ & 0.44 & 0.35 & 0.47 & 0.05 & 6.84 & 16.66 & 3.38 
		& 0.13 & 0.60 & 0.28 \\
		& $\sigma_{x}/\overline{x}$ & 25 \% & 31 \% & 6 \% & 6 \% & 14 \% & 
		11 \% & 4 \% & 6 \% & 6 \% & 5 \% \\
		\cline{2-12} \cline{2-12} & $\overline{x}$ & 1.82 & 1.22 & 7.92 & 
		0.76 & 46.96 & 155.70 & 93.54 & 2.30 & 9.49 & 6.74 \\
		G & $\sigma_{x}$ & 0.39 & 0.25 & 2.08 & 0.28 & 6.52 & 14.48 & 2.90 
		& 0.29 & 0.44 & 2.23 \\
		& $\sigma_{x}/\overline{x}$ & 21 \% & 20 \% & 26 \% & 37 \% & 14 \% 
		& 9 \% & 3 \% & 13 \% & 5 \% & 33 \% \\
		\cline{2-12} \cline{2-12} & $\overline{x}$ & 2.59 & 0.55 & 5.38 & 
		0.72 & 56.68 & 140.80 & 98.68 & 2.23 & 9.43 & 5.82 \\
		U & $\sigma_{x}$ & 0.11 & 0.06 & 0.37 & 0.11 & 4.31 & 11.68 & 3.00 
		& 0.20 & 0.22 & 0.19 \\
		& $\sigma_{x}/\overline{x}$ & 4 \% & 10 \% & 7 \% & 15 \% & 8 \% & 
		8 \% & 3 \% & 9 \% & 2 \% & 3 \% \\
		\cline{2-12}
	\end{tabular}
	\caption{PCP of amino-acids coded by codons with same second nucleotide 
	(written at the left)}
	\label{tableNC}
\end{table}

\clearpage

\begin{table}
	\footnotesize
	\centering
	\begin{tabular}{l|c||c|c|c|c|c|c|c|c|c|c|}
		\cline{2-12} & & $P_{\alpha} + P_{\beta}$ & $P_{\tau}$ & $P_{G}$ & 
		$R_{f}$ & $\Delta H$ & $-\Delta S$ & $\Delta G$ & $ pK_{a}$ & 
		$pK_{b}$ & $pI$ \\
		\cline{2-12} \cline{2-12} & $\overline{x}$ & 2.04 & 0.97 & 10.45 & 
		0.83 & 61.10 & 121.80 & 95.95 & 2.00 & 9.15 & 6.56 \\
		Gln/His & $\sigma_{x}$ & 0.17 & 0.01 & 0.05 & 0.01 & $-$ & $-$ & 
		1.36 & 0.17 & 0.02 & 0.91 \\
		& $\sigma_{x}/\overline{x}$ & 8 \% & 2 \% & 0 \% & 1 \% & $-$ & $-$ 
		& 1 \% & 9 \% & 0 \% & 14 \% \\
		\cline{2-12} \cline{2-12} & $\overline{x}$ & 1.72 & 1.10 & 12.60 & 
		0.86 & & & & 2.04 & 9.64 & 3.00 \\
		Asp/Glu & $\sigma_{x}$ & 0.17 & 0.36 & 0.40 & 0.02 & & & & 0.16 & 
		0.04 & 0.22 \\
		& $\sigma_{x}/\overline{x}$ & 10 \% & 33 \% & 3 \% & 2 \% & & & & 8 
		\% & 0 \% & 8 \% \\
		\cline{2-12} \cline{2-12} & $\overline{x}$ & 1.75 & 1.29 & 11.40 & 
		0.93 & 56.40 & 135.70 & 96.84 & 2.10 & 8.88 & 7.50 \\
		Asn/Lys & $\sigma_{x}$ & 0.15 & 0.28 & 0.10 & 0.04 & 1.40 & 0.10 & 
		1.32 & 0.08 & 0.08 & 2.09 \\
		& $\sigma_{x}/\overline{x}$ & 9 \% & 21 \% & 1 \% & 4 \% & 2 \% & 0 
		\% & 1 \% & 4 \% & 1 \% & 28 \% \\
		\cline{2-12} \cline{2-12} & $\overline{x}$ & 2.03 & 1.17 & 5.85 & 
		0.67 & 49.15 & 151.10 & 94.29 & 2.08 & 9.70 & 5.34 \\
		Cys/Tyr & $\sigma_{x}$ & 0.14 & 0.02 & 0.35 & 0.18 & 5.25 & 12.10 & 
		1.40 & 0.12 & 0.58 & 0.32 \\
		& $\sigma_{x}/\overline{x}$ & 7 \% & 2 \% & 6 \% & 27 \% & 11 \% & 
		8 \% & 1 \% & 6 \% & 6 \% & 6 \% \\
		\cline{2-12} \cline{2-12} & $\overline{x}$ & 1.32 & 1.48 & 8.60 & 
		0.89 & 43.35 & 172.25 & 94.74 & 2.10 & 9.88 & 5.99 \\
		Pro/Ser & $\sigma_{x}$ & 0.20 & 0.04 & 0.60 & 0.07 & 7.35 & 7.75 & 
		4.76 & 0.11 & 0.73 & 0.31 \\
		& $\sigma_{x}/\overline{x}$ & 15 \% & 3 \% & 7 \% & 8 \% & 17 \% & 
		4 \% & 5 \% & 5 \% & 7 \% & 5 \% \\
		\cline{2-12} \cline{2-12} & $\overline{x}$ & 2.14 & 0.81 & 8.34 & 
		0.91 & 52.05 & 141.9 & 94.32 & 2.22 & 9.40 & 5.82 \\
		Ala/Thr & $\sigma_{x}$ & 0.12 & 0.15 & 0.25 & 0.02 & 1.35 & 5.90 & 
		0.34 & 0.13 & 0.29 & 0.18 \\
		& $\sigma_{x}/\overline{x}$ & 5 \% & 19 \% & 3 \% & 2 \% & 3 \% & 4 
		\% & 0 \% & 6 \% & 3 \% & 3 \% \\
		\cline{2-12} \cline{2-12} & $\overline{x}$ & 2.59 & 0.54 & 5.45 & 
		0.75 & 57.75 & 140.40 & 99.63 & 2.32 & 9.41 & 5.84 \\
		Ile/Met & $\sigma_{x}$ & 0.09 & 0.07 & 0.25 & 0.01 & 0.45 & 11.90 & 
		2.97 & 0.04 & 0.19 & 0.10 \\
		& $\sigma_{x}/\overline{x}$ & 3 \% & 12 \% & 5 \% & 1 \% & 1 \% & 8 
		\% & 3 \% & 2 \% & 2 \% & 2 \% \\
		\cline{2-12} \cline{2-12} & $\overline{x}$ & 2.64 & 0.55 & 5.40 & 
		0.79 & 59.80 & 133.55 & 99.70 & 2.34 & 9.61 & 5.97 \\
		Leu/Val & $\sigma_{x}$ & 0.12 & 0.05 & 0.50 & 0.06 & 0.40 & 5.45 & 
		2.00 & 0.02 & 0.01 & 0.01 \\
		& $\sigma_{x}/\overline{x}$ & 5 \% & 8 \% & 9 \% & 8 \% & 1 \% & 4 
		\% & 2 \% & 1 \% & 0 \% & 0 \% \\
		\cline{2-12} \cline{2-12}
	\end{tabular}
	\caption{Table of amino-acid doublet properties} \label{tableSD}
\end{table}

\end{document}